\documentstyle[11pt]{article}
\input amssymb.sty

\begin{document}

\newcommand{\Si}{\Sigma}
\newcommand{\tr}{{\rm tr}}
\newcommand{\ad}{{\rm ad}}
\newcommand{\Ad}{{\rm Ad}}
\newcommand{\de}{\delta}
\newcommand{\al}{\alpha}
\newcommand{\te}{\theta}
\newcommand{\vth}{\vartheta}
\newcommand{\la}{\lambda}
\newcommand{\La}{\Lambda}
\newcommand{\D}{\Delta}
\newcommand{\ve}{\varepsilon}
\newcommand{\ep}{\epsilon}
\newcommand{\vf}{\varphi}
\newcommand{\G}{\Gamma}
\newcommand{\ka}{\kappa}
\newcommand{\ip}{\hat{\upsilon}}
\newcommand{\Ip}{\hat{\Upsilon}}
\newcommand{\ga}{\gamma}
\newcommand{\ze}{\zeta}
\newcommand{\si}{\sigma}
\newcommand{\na}{\nabla}
\newcommand{\om}{\omega}
\newcommand{\Om}{\Omega}

\def\we{\wedge}

\def\mC{{\mathbb C}}
\def\mZ{{\mathbb Z}}
\def\mR{{\mathbb R}}
\def\mN{{\mathbb N}}

\def\frak{\mathfrak}
\def\gg{{\frak g}}
\def\gJ{{\frak J}}
\def\gS{{\frak S}}
\def\gL{{\frak L}}
\def\gG{{\frak G}}
\def\gk{{\frak k}}
\def\gK{{\frak K}}
\def\gl{{\frak l}}
\def\gh{{\frak h}}

\def\bfa{{\bf a}}
\def\bfb{{\bf b}}
\def\bfc{{\bf c}}
\def\bfd{{\bf d}}
\def\bfe{{\bf e}}
\def\bfm{{\bf m}}
\def\bfn{{\bf n}}
\def\bfp{{\bf p}}
\def\bfu{{\bf u}}
\def\bfr{{\bf r}}
\def\bfv{{\bf v}}
\def\bft{{\bf t}}
\def\bfx{{\bf x}}
\def\bfg{{\bf g}}
\def\bfM{{\bf M}}
\def\bfnu{{\bf \nu}}
\def\bfS{{\bf S}}
\def\bfJ{{\bf J}}
\def\bfC{{\bf C}}
\newcommand{\li}{\lim_{n\rightarrow \infty}}
\newcommand{\mat}[4]{\left(\begin{array}{cc}{#1}&{#2}\\{#3}&{#4}
\end{array}\right)}
\newcommand{\thmat}[9]{\left(
\begin{array}{ccc}{#1}&{#2}&{#3}\\{#4}&{#5}&{#6}\\
{#7}&{#8}&{#9}
\end{array}\right)}
\newcommand{\beq}[1]{\begin{equation}\label{#1}}
\def\eq{\end{equation}}
\newcommand{\beqn}[1]{\begin{eqnarray}\label{#1}}
\newcommand{\eqn}{\end{eqnarray}}
\newcommand{\pa}{\partial}
\newcommand{\di}{{\rm diag}}
\newcommand{\ti}{\tilde}
\newcommand{\oh}{\frac{1}{2}}

\newcommand{\su}{{\bf su_2}}
\newcommand{\uo}{{\bf u_1}}
\newcommand{\GL}{{\rm GL}(N,\mC)}
\newcommand{\SLN}{{\rm SL}(N,\mC\,)}
\newcommand{\SLt}{{\rm SL}(3,\mC)}
\def\sln{{\rm sl}(N,\mC)}
\def\gln{{\rm gl}(N,\mC)}
\newcommand{\PSL}{{\rm PSL}_2({\mZ})}
\def\SL2{{\rm SL}(2,\mC)}

\newcommand{\ran}{\rangle}
\newcommand{\lan}{\langle}
\def\f1#1{\frac{1}{#1}}
\def\lb{\lfloor}
\def\rb{\rfloor}
\newcommand{\rar}{\rightarrow}
\newcommand{\upar}{\uparrow}
\newcommand{\sm}{\setminus}
\newcommand{\ms}{\mapsto}
\newcommand{\bp}{\bar{\partial}}
\newcommand{\bz}{\bar{z}}
\newcommand{\bA}{\bar{A}}
\newcommand{\sect}[1]{\setcounter{equation}{0}\section{#1}}
\renewcommand{\theequation}{\thesection.\arabic{equation}}
\newtheorem{predl}{Proposition}[section]
\newtheorem{defi}{Definition}[section]
\newtheorem{rem}{Remark}[section]
\newtheorem{cor}{Corollary}[section]
\newtheorem{lem}{Lemma}[section]
\newtheorem{theor}{Theorem}[section]

\vspace{0.3in}
\begin{flushright}
 ITEP-TH-40/04\\
\end{flushright}
\vspace{10mm}
\begin{center}
{\Large\bf
Universality of Calogero-Moser model}\\
\vspace{5mm}

M.A.Olshanetsky
\\
{\sf RIMS, Kyoto University, Japan;}\\
{\sf Institute of Theoretical and Experimental Physics, Moscow, Russia,}\\
{\em e-mail olshanet@itep.ru}\\

\vspace{5mm}
\end{center}

\begin{abstract}
\noindent
In this review we explain interrelations between the Elliptic Calogero-Moser
model, integrable Elliptic Euler-Arnold top, monodromy preserving
equations and the Knizhnik-Zamolodchikov-Bernard equation on a torus.

\end{abstract}

\begin{flushright}
 {\sl Dedicated to 70-th birthday of Francesco Calogero}\\
\end{flushright}

\tableofcontents


\section {Introduction}
\setcounter{equation}{0}

The Calogero Model first proposed by Francesco Calogero as a model of exactly
solvable one-dimensional nuclei \cite{Ca1,Ca2}.
Later different generalizations of the model on the classical and quantum
level were introduced in Refs.\,\cite{M,OP1,S} (see also reviews \cite{OP2,OP3}).
Nowadays these models, that we will call for brevity the Calogero model
(CM), play a fundamental role in the contemporary theoretical physics.
We shortly remind some of them. The first indication of this role came from
the papers \cite{AMM,Kr} where interrelations between
  classical solutions of the rational and
elliptic CM and special solutions of the KdV and KP
equations were established.
 Last fifteen years a wide range of applications was discovered.
Among them are
interrelations between the Calogero-Sutherland model \cite{S} and the
Fractional Quantum Hall Effect \cite{AI}, integrable one-dimensional spin models
with long-range interactions \cite{Ino}. Important role plays the classical
CM in the SUSY Yang-Mills theory \cite{GKMMM} and in the string theory \cite{V}.

Most likely, the fundamental character of CM can be explained by their group-theoretical
and geometrical nature. In the very beginning of seventies during Francesco
Calogero visit to ITEP Ascold Perelomov and I have realized that the
Calogero-Sutherland Hamiltonians up to a conjugation coincide with the
radial parts of the second Casimir operators on $\sln$ and $\SLN$.
This observation was a
starting point of our investigations of classical and quantum integrable systems,
related to Lie algebras. According to this approach it was established that solutions of
the classical rational and the trigonometric models
come from a free motion on Lie algebras and Lie groups \cite{KKS,OP3,OP4}.
In this way their quantum counterparts are related to the representation
theory of simple Lie algebras \cite{OP5}. It imlies, in particular, that the wave-functions are just
some special matrix elements in irreducible representations.

In the elliptic case the situation
is more elaborate. The classical elliptic Calogero-Moser model (ECMM) is a particular
example of the Hitchin systems \cite{H}. It is a wide class of classical
integrable systems that come from a topological 3d gauge theory. The
inclusion of CM in the Hitchin theory was observed independently in Refs.\,\cite{ER,GN,Mar,N}.

In this brief review we touch another facets of the classical and
quantum ECCM. In Sect.1 we discuss
equivalence of the classical ECMM and the
so-called elliptic top (ET). The later describes the classical degrees of freedom
that located on a vertex of the $\SLN$ generalization of the XYZ lattice model.
For the two-particle case it leads to the equivalence between the
two-dimensional version of ECMM \cite{Kr1,LOZ1} and the Landau-Lifshitz model. This section
is based on Ref.\,\cite{LOZ1}. The correspondence between the classical ECMM
for two particle case and the  ${\rm Painlev\grave{e}}$ VI equation \cite{LO} is discussed in
Sect.2. Finally, in Sect.3 we present the interpretation of the
${\rm Schr\ddot{o}dinger}$ equation corresponding to the
quantum ECMM and  Knizhnik-Zamolodchikov-Bernard equation that arises in
the Wess-Zumino-Witten model on a torus.

\section{Calogero-Moser model and Integrable tops}
\setcounter{equation}{0}

\subsection{ECMM with spin}
{\sl Description of the system}

The ECMM system is defined by the Hamiltonian
\beq{1.1}
H^{CM}=\oh\sum_{j=1}^Nv_j^2+\nu^2\sum_{j>k}\wp(u_j-u_k;\tau)
\end{equation}
on the phase space $\mC^{2N}$ with  the canonical brackets
\beq{2.1}
\{v_j,u_k\}=\de_{j,k}\,.
\end{equation}
Here $\bfu=(u_1,u_2,\ldots,u_N)$ are coordinates and $\bfv=(v_1,v_2,\ldots,v_N)$
are their momenta. In what follows we assume that $\sum_ju_j=0$,
$\sum_jv_j=0$.

Let $T^2_\tau=\mC\,/(\mZ\oplus\tau\mZ)$
be a torus endowed with a complex structure with  parameter
$\tau\,,\Im m\tau>0$.
The double-periodicity
of the Weierstrass function implies that the particles lie on the torus
$u_j\in T^2_\tau$, while $\bfv\in\mC^N$.
In fact, in the potential in (\ref{1.1}) we will consider another
double-periodic function
$$
E_2(x;\tau)=\wp(x;\tau)+2\eta_1(\tau)\,,
$$
where $E_2$ is the second Eisenstein function and $\eta_1(\tau)=\zeta(\oh)$.
The additional constant becomes essential only on the quantum level.

The system has the "spin" generalization \cite{GH}. Let $\bfp$ be an $N$-order
matrix.
 We consider $\bfp$ as an
element of the Lie algebra $\sln$. The linear (Lie-Poisson) brackets on
$\sln$  for the matrix elements assume the form
\beq{3.1}
\{p_{jk},p_{mn}\}=p_{jn}\de_{km}-p_{mn}\de_{jk}\,.
\end{equation}
Let ${\cal O}$ be a coadjoint orbit
\beq{orb}
{\cal O}=\{p\in\sln~|~p=h^{-1}p^0h,~h\in\SLN\,,p^0\in D\}\,,
\end{equation}
where $D$ is the diagonal subgroup of $\SLN$.
The phase space of the ECMM with spin is
\beq{b15}
{\mathcal R}^{CM_N} =\{T^*_\mC(T^2_\tau)^{N-1},\ti{\cal O}\}\,,
\end{equation}
where $\ti{\cal O}={\cal O}//D$ is the symplectic quotient
with respect to the action of $D$. It implies  i) the moment constraint
$p_{jj}=0$,
ii) the gauge fixing, for example, as $p_{j,j+1}=p_{j+1,j}$.\\
Note that
\beq{od}
\dim({\mathcal R}^{CM_N})=2N-2+\dim{\cal O}-2\dim(D)=\dim{\cal O}\,.
\end{equation}

The spin ECMM Hamiltonian has the form
\beq{4.1}
H^{CM,spin}=\oh\sum_{j=1}^Nv_j^2+\sum_{j>k}p_{jk}p_{kj}E_2(u_j-u_k;\tau)\,.
\end{equation}
The case (\ref{1.1}) corresponds to the most degenerate nontrivial orbit
${\cal O}\,\sim\, T^*\mC P^{N-1}$ when $N-1$ eigen-values coincide.
In this case $\dim(\ti{\cal O})=0$ . The coupling constant $\nu^2$ is proportional
to  $\tr(\bfp^2)$.

The equations of motion can be read off from (\ref{2.1}), (\ref{3.1}) and
(\ref{4.1})
\beq{5.1}
\pa_tu_j=v_j\,,
\end{equation}
\beq{6.1}
\pa_tv_n=-\sum_{j\neq n}p_{jk}p_{kj}\pa_{u_n}E_2(u_j-u_n;\tau)\,,
\end{equation}
\beq{7.1}
\pa_t\bfp=2[\bfp,\bfJ_\bfu (\bfp)]\,,
\end{equation}
where the operator $\bfJ_\bfu$ acts on $\sln$ as
$
\bfJ_\bfu\,:~p_{jk}\,\to\,E_2(u_j-u_k)\pa_{jk}.
$

\bigskip
{\sl Lax representation}

The system has the Lax representation
$$
\pa_tL^{CM}=[L^{CM},M^{CM}]\,,
$$
Introduce an auxiliary elliptic curve $E_\tau$ with the same modular parameter as above.
It plays the role of the basic spectral curve with the
spectral parameter $z$. The Lax matrix depends on $z$
and has the form
\beq{l1}
L^{CM}=P+X,~~P=\di(v_1,\ldots,v_N),~~X_{jk}=p_{jk}\phi(u_j-u_k,z)\,,
\end{equation}
where
\beq{A.3}
\phi(u,z)=
\frac
{\vth(u+z)\vth'(0)}
{\vth(u)\vth(z)}\,,
\end{equation}
and $\vth(z)=\vth(z|\tau)$ is the odd theta-function
$$
\vth(z|\tau)=q^{\frac
{1}{8}}\sum_{n\in {\bf Z}}(-1)^ne^{\pi i(n(n+1)\tau+2nz)}\,,
~~(q=\bfe(\tau)=\exp 2\pi i\tau)\,.
$$
The matrix $M^{CM}$ corresponding to the flow (\ref{5.1})--(\ref{7.1}) takes
the form
\beq{l2}
M^{CM}=-D+Y\,,~~D=\di(Z_1,\ldots,Z_N)\,,~~Y_{jk}=y(u_j-u_k,z)\,,
\end{equation}
$$
Z_j=\sum_{k\neq j}E_2(u_j-u_k)\,,~~y(u,z)=\frac{\pa \phi(u,z)}{\pa u}\,.
$$

The Lax matrix is a quasi-periodic meromorphic functions on the spectral
curve $E_\tau$ taking values in the Lie algebra $\sln$
 with a simple pole at $z=0$
\beq{20.1}
\bp L^{CM}=0\,,~~Res L^{CM} |_{z=0}=\bfp\,.
\end{equation}
\beq{23.1}
L^{CM}(z+1)=L^{CM}(z)\,,~~L^{CM}(z+\tau)=\di(\bfe(\bfu))L^{CM}(z)\di(-\bfe(\bfu))\,,
\end{equation}
where $\di(\bfe(\bfu))=\di(\exp(2\pi iu_1,\ldots,\exp(2\pi iu_N))$.
These conditions uniquely characterized the non-diagonal part $X$ of $L^{CM}$.

The Lax equation is equivalent to the linear problem
\beq{L1}
(\la+L^{CM})Y=0\,,
\end{equation}
\beq{L2}
\pa_t+M^{CM})Y=0\,.
\end{equation}
The additional equation
\beq{L3}
\bp Y=0
\end{equation}
implies that  $M^{CM}$ is also meromorphic on $E_\tau$.


\subsection{Elliptic top on $\SLN$}

{\sl Description of the top}

Consider the Euler-Arnold top (EAT) on the group $\SLN$. Its phase
space is embedded in the Lie coalgebra $\sln^*$  as a coadjoint orbit.
It is endowed with Lie-Poisson brackets (\ref{3.1}).

The EAT is determined by a symmetric operator
$\bfJ\,:\,\sln^*\to\sln$,
that is called the inverse inertia operator.
The Hamiltonian of the system is $H^{EAT}=\tr(\bfS\bfJ(\bfS))$, where
$\bfS\in\sln^*$. A special choice of $\bfJ$ leads to an integrable system.
The elliptic top (ET) is an example of an integrable EAT.

To define the inverse inertia operator for ET we
choose another basis in $\sln^*\sim\sln$. Define two type of matrices
$$
Q=\di({\bf e}_N(1),\ldots,{\bf e}_N(m),\ldots,1)\,,
~(\,{\bf e}_N(z)=\exp (\frac{2\pi i}{N} z)\,)\,,
$$
$$
\La=
\left(\begin{array}{ccccc}
0&1&0&\cdots&0\\
0&0&1&\cdots&0\\
\vdots&\vdots&\ddots&\ddots&\vdots\\
0&0&0&\cdots&1\\
1&0&0&\cdots&0
\end{array}\right)\,.
$$

Consider a two-dimensional lattice of order $N^2-1$
$ \mZ^{(2)}_N=(\mZ/N\mZ\oplus\mZ/N\mZ)/(0,0)$.
The matrices
$$
T_{\al}=\f1{2\pi i\te}\bfe_N(\frac{\al_1\al_2}{2})Q^{\al_1}\La^{\al_2}\,,
~(\al=(\al_1,\al_2)\in\mZ^{(2)_N})
$$
generate a basis in $\sln$.
 The commutation relations in this basis assume the form
$$
[T_{\al},T_{\beta}]=\bfC_N(\al,\beta)T_{\al+\beta}\,,~~
$$
where
\beq{AA4}
\bfC_N(\al,\beta)=\frac{N}{\pi}\sin\frac{\pi(\al\times \beta)}{N}\,.
\end{equation}

 The the Poisson structure on the dual space
$\sln^*$
is given by the linear Lie-Poisson brackets coming from (\ref{AA4})
\beq{lb}
\{S_\al,S_\beta\}_1=\bfC(\al,\beta)S_{\al+\beta}\,.
\end{equation}

Let $\mZ_N^{(2)}(\tau)=\frac{\ga_1+\ga_2\tau}{N}$, $\ga\in\mZ^{(2)}_N$
be a regular lattice of order $N^2-1$ on $T^2_\tau$.
Introduce the following  constant on $\mZ_N^{(2)}(\tau)$:
$E_2(\ga)=E_2(\frac{\ga_1+\ga_2\tau}{N})$.
Then the operator $\bfJ$ for the ET is defined as
\beq{40.1}
\bfJ\,:\, S_\al\,\to\,E_2(\al)S_\al\,.
\end{equation}

 Let $\bfS=\sum_{\al\in \mZ^{(2)}_N} S_{-\al} T_\al$.
 The Hamiltonian has the form
\beq{8.5}
H^{ET}=-\oh\tr(\bfS\cdot \bfJ(\bfS))
\equiv -\oh\sum_{\ga\in \mZ^{(2)}_N} S_\ga E_2(\ga) S_{-\ga}\,.
\end{equation}
It defines the equations of motion
\beq{em}
\pa_t\bfS=[\bfJ(\bfS),\bfS]\,,
\end{equation}
or
\beq{8.5a}
\pa_tS_\al=\sum_{\ga\in \mZ^{(2)}_N}
 S_{\al-\ga}S_\ga E_2(\ga)\bfC_\te(\ga,\al)\,.
\end{equation}
The phase space ${\cal R}^{ET}$ of ET is a coadjoint orbit of $\SLN$
\beq{eto}
{\cal R}^{ET}={\cal O}\,.
\end{equation}
Note that it dimension coincides with $\dim({\cal R}^{CM})$.

The Lax form of (\ref{em}) is provided by the Lax matrix
\cite{STS}
\beq{8.6}
L^{ET}=\sum_\al S_\al\vf(\al,z)T_\al\,,
~~\vf(\ga,z)=\bfe(\frac{\ga_2z}{N})\phi(\frac{\ga_1+\ga_2\tau}{N},z)\,,
\end{equation}
 and
\beq{8.7}
M^{ET}=\sum_\al S_\al f(\al,z)T_\al\,,~~
f(\al,z)=\bfe(\frac{\al_2z}{N})\partial_{u}\phi(u;z)|_{u=\frac{\al_1+\al_2\tau}{N}}\,.
\end{equation}
The Lax matrix is characterized by the following conditions:
\beq{22.1}
\bp L^{ET}=0\,,~~Res L^{ET} |_{z=0}=\bfS=\sum S_\al T_\al\,,
\end{equation}
\beq{24.1}
L^{ET}(z+1)=Q(\tau)L^{ET}(z)Q^{-1}(\tau)\,,
~
L^{ET}(z+\tau)=\ti{\La}(z,\tau)L^{ET}(z)(\ti{\La}(z,\tau))^{-1}\,,
\end{equation}
where $\ti{\La}(z,\tau)=-\bfe(\frac{-z-\oh\tau}{N})\La$.


\subsection{The map of the ECMM system to the ET system}

The  map is defined as the conjugation of $L^{CM}$ by
some matrix $\Xi(z)$:
\beq{8.11a}
L^{ET}=\Xi\times L^{CM}\times \Xi^{-1}\,.
\end{equation}
The matrix $\Xi(z)$ is a meromorphic quasi-periodic map $E_\tau\to\GL$.
It is uniquely  defined by its quasi-periodicity and
the pole at $z=0$.
The latter means that $\Xi$
 can be considered as a singular
gauge transformation. Assume that an eigen-vector
of the
residue of $L^{CM_N}=\bfp$ at $z=0$ belongs to the kernel of $\Xi(z)$. Then it
can be proved that (\ref{8.11a}) preserves the order of the pole.

The matrix $\Xi$ has the following form.
 The quasi-periodicity of $L^{CM}$ and $L^{ET}$ leads to the following
relations
\beq{8.12}
\Xi(z+1,\tau)= Q\times \Xi(z,\tau)\,,
\end{equation}
\beq{8.13}
\Xi(z+\tau,\tau)=\tilde\Lambda(z,\tau)\times
\Xi(z,\tau)
\times{\rm diag}
({\bf e}(u_j))\,.
\end{equation}

Let $\bfp^0$ be the diagonal matrix defining the coadjoint
orbit (\ref{orb}) in the ECMM
\beq{p}
Res\, L^{CM_N}|_{z=0}=\bfp=h^{-1}\bfp^0h\,,~~
\bfp^0=\di(p^0_1,\ldots,p^0_N)\,.
\end{equation}
 Then $\Xi(z)=\Xi(z,\overrightarrow{\bfr}_j)$ depends
on a choice of the eigen-vector
 $\overrightarrow{\bfr}_j=(r_{1,j},\ldots,r_{N,j})$
 of the orbit matrix $\bfp$, that belongs to the kernel of $\Xi$ and
corresponds to the eigenvalue $p_j^0$ (\ref{p}). It has the form
$\overrightarrow{\bfr}_j=h^{-1}(0,\ldots,0,1,0\ldots,0)^T$,
where $1$ stands on the $j$-th place and $h$ is defined up to a maximal parabolic subgroup.

We construct first $(N\times N)$-
matrix $\tilde\Xi(z, u_1,\ldots,u_N;\tau)$ that satisfies (\ref{8.12}) and
(\ref{8.13}) but has a special kernel:
\beq{8.14}
\tilde\Xi_{ij}(z, u_1,\ldots,u_N;\tau) =
\theta{\left[\begin{array}{c}
\frac{i}N-\frac12\\
\frac{N}2
\end{array}
\right]}(z-Nu_j, N\tau ),
\end{equation}
where $\theta{\left[\begin{array}{c}
a\\
b\end{array}
\right]}(z, \tau )$ is the theta function with a characteristic
$$
\theta{\left[\begin{array}{c}
a\\
b
\end{array}
\right]}(z , \tau )
=\sum_{j\in \Bbb Z}
{\bf e}\left((j+a)^2\frac\tau2+(j+a)(z+b)\right)\,.
$$

It can be proved that the kernel of $\tilde\Xi$ at $z=0$ is generated
by the following column-vector~:
$$
\left\{(-1)^l \prod_{j<k;j,k\ne l}
\vartheta(u_k-u_j,\tau)\right\}, \quad l=1,2,\cdots,N.
$$

Then the matrix $\Xi(z,\bfu,\overrightarrow{\bfr}_j)$ assumes the form
\beq{xi1}
\Xi(z,\bfu,\overrightarrow{\bfr}_j)=\tilde{\Xi}(z)\times{\rm diag}
\left(\frac{(-1)^l}{r_{l,j}}
\prod_{j<k;j,k\ne l}
\vartheta(u_k-u_j,\tau)\right).
\end{equation}
It leads to the map
${\mathcal R}^{CM_N}\to {\mathcal R}^{rot}$.

This transformation means that the particle degrees of freedom of the ECMM
$(\bfv,\bfu)$ along with the spin variables $\bfp$ boil down to
the orbits variables $\bfS$ $(\bfv,\bfu,\bfp)\mapsto \bfS$.
 For the most degenerate orbit in the standard ECMM, defined by the coupling
constant $\nu^2$ this transformation leads to the degenerate orbit of the ET
with the same value of Casimir. Note that equation for ECMM with spin
(\ref{6.1}) remind the equation of motion for the EAT
 with the time-dependent operator $\bfJ_\bfu$ (\ref{4.1}), (\ref{5.1}).
The only difference is the structure of the phase spaces
${\mathcal R}^{CM_N}$ (\ref{b15}) and ${\cal R}^{ET}$ (\ref{eto}). The gauge
transform $\Xi$ carries out the pass from ${\mathcal R}^{CM_N}$ to
${\cal R}^{ET}$. It depends only on the part of variablis on
${\mathcal R}^{CM_N}$, namely on $\bfu$ and $\bfp$ through the eigenvector
$\overrightarrow{\bfr}_j$.

Consider in detail the case $N=2$, when the system has the one degree of freedom.
 Let the eigen-vector of $\bfp$ has the form $(1,1)^T$ and put
$\bfS=S_a\si_a$,
where $\si_a$ denote the sigma matrices.
Then the transformation has the form
\beq{S}
\left\{
\begin{array}{l}
S_1=-v\frac{\theta_{10}(0)}{\vartheta'(0)}
\frac{\theta_{10}(2u)}{\vartheta(2u)}-
\nu\frac{\theta_{10}^2(0)}{\theta_{00}(0)\theta_{01}(0)}
\frac{\theta_{00}(2u)\theta_{01}(2u)}{\vartheta^2(2u)}\,,
 \\
S_2=-v\frac{\theta_{00}(0)}{i\vartheta'(0)}
\frac{\theta_{00}(2u)}{\vartheta(2u)}-
\nu\frac{\theta_{00}^2(0)}{i\theta_{10}(0)\theta_{01}(0)}
\frac{\theta_{10}(2u)\theta_{01}(2u)}{\vartheta^2(2u)}\,,
\\
S_3=-v\frac{\theta_{01}(0)}{\vartheta'(0)}
\frac{\theta_{01}(2u)}{\vartheta(2u)}-
\nu\frac{\theta_{01}^2(0)}{\theta_{00}(0)\theta_{10}(0)}
\frac{\theta_{00}(2u)\theta_{10}(2u)}{\vartheta^2(2u)}\,.
 \\
\end{array}
\right.
\end{equation}


\section{Calogero-Moser model and Isomonodromic deformations}
\setcounter{equation}{0}

The famous  Painlev\'{e} VI
equation depends on four free parameters
(PVI$_{\al,\beta,\ga,\de}$) and has the
form
$$
\frac{d^2X}{dt^2}=\frac{1}{2}\left(\frac{1}{X}+\frac{1}{X-1}+
\frac{1}{X-t}\right)
\left(\frac{dX}{dt}\right)^2-
\left(\frac{1}{t}+\frac{1}{t-1}+\frac{1}{X-t}\right)\frac{dX}{dt}+
$$
\beq{1.3}
+\frac{X(X-1)(X-t)}{t^2(t-1)^2}\left(\al+\beta\frac{t}{X^2}+
\ga\frac{t-1}{(X-1)^2}
+\de\frac{t(t-1)}{(X-t)^2}\right).
\end{equation}

It can be transformed to the elliptic form \cite{BB,M,P} that we will use.
Let $\om_0=0$, $\om_j$ are the half periods of the elliptic curve
$E_\tau$ and
$$
\nu_0=\al\,,~\nu_1=-\beta\,,~\nu_2=\ga\,,~\nu_3=\oh-\de\,.
$$
Then (\ref{1.3}) takes the form
\beq{fp}
\pa_\tau u=-\sum_{j=0}^3\pa_u\nu_j^2E_2(u+\om_j)\,,
\end{equation}
wherethe variables are  replaced  as
$$
(u,\tau)\rar
\left(X=\frac{E_2(u|\tau)-e_1}{e_2-e_1}\,,~
Y=\frac{\pa_u E_2(u|\tau)}{e_2-e_1}\,,~
t=\frac{e_3-e_1}{e_2-e_1}\right)\,,~e_j=E_2(\om_j)\,.
$$

We will not consider here the general case and restrict ourselves to the case
$\nu_j=\nu$
\footnote{the general case was investigated in \cite{Z}.}.
Then PVI assumes the form
$$
\pa^2_\tau u=-\pa_u\nu^2E_2(2u)\,.
$$
It is a non-autonomous Hamiltonian system with the same Hamiltonian as for
the two-body ECMM
$$
H=\oh v^2+\nu^2E_2(2u)\,,
$$
but now the module $\tau$ plays the role of the time.

Let us introduce the new parameter $\ka$ and consider the equation
\beq{I.3a}
\ka^2\frac{d^2u}{d\tau^2}=-\pa_u\nu^2E_2(2u|\tau).
\end{equation}
It can be achieved by the rescaling the dynamical variables $(v,u)$ and the
half-periods $\om_1,\om_2$
\beq{hp}
v\rar\ka^{-\oh}v\,,~u\rar\ka^{\oh}u\,,~\om_j\rar\ka^{\oh}\om_j\,.
\end{equation}

The equation (\ref{I.3a}) has the Lax representation
$$
\pa_\tau L^P-\ka\pa M^P+[M^P,L^P]=0\,.
$$
Let $\mu=\frac{\tau-\tau_0}{\rho}$, $\rho=\tau_0-\overline{\tau}_0$,
$x(u,w,\bar{w})=\frac{\nu}{2\pi i(1-\mu)}\phi(u,w)$, where
$\tau_0$ corresponds to some fixed module and
  $\phi(u,w)$ is deined by (\ref{A.3}),  $y(u,w,\bar{w})=
\frac{\rho}{2\pi i\ka(\tau-\bar{\tau}_0)}\pa_u x(u,w,\bar{w})$.
The Lax matrices assume the
form
\beq{3.10}
L^P= \mat{\frac{v}{1-\ti{\mu}_\tau}-\ka\frac{u}{\rho}}
{x(u,w,\bar{w})}{x(-u,w,\bar{w})}
{-\frac{v}{1-\ti{\mu}_\tau}+\ka\frac{u}{\rho}}\,,~
~
M^P=\mat{0}{y(2u,w,\bar{w})}{y(-2u,w,\bar{w})}{0},
\end{equation}
The Lax equation can be
considered as the consistency condition for the linear system
\beq{ls4}
(\pa+L^P)\Psi=0\,,
\end{equation}
\beq{ls6}
(\ka\pa_\tau+M^P)\Psi=0\,,
\end{equation}
\beq{ls5}
(\bp+\mu\pa)\Psi=0\,,~~\mu=\frac{\tau-\tau_0}{\tau_0-\overline{\tau}_0}
\end{equation}
where
(\ref{ls5}) implies the holomorphity of the Baker-Akhiezer
function $\Psi$ in the coordinates deformed by
$\mu$: $w=z-\mu(\bz-z)$, $\bar{w}=\bz$.

 We will prove that the
linear problem for the two-body ECMM (\ref{L1})--(\ref{L3})  coincides with (\ref{ls4})--(\ref{ls6})
in the limit $\ka\to 0$. The constant $\ka$ plays the role of the Planck
constant and  (\ref{L1})--(\ref{L3}) is the result of the quasi-classical limit.
Define the time $t$ corresponding to the two-body ECMM Hamiltonian as
$\tau=\tau_0+\ka t$,
and represent the Baker-Akhiezer function in the WKB approximation form
\beq{WKB}
\Psi=\Phi\exp(\frac{{\cal S}^{(0)}}{\ka}+{\cal S}^{(1)})\,,
\end{equation}
where $\Phi$ is a  group valued function and ${\cal S}^{(0)}$,
${\cal S}^{(1)}$ are diagonal matrices. Substitute (\ref{WKB})
 in the linear system (\ref{ls4}),(\ref{ls5}),(\ref{ls6}). If
$\frac{\pa}{\pa\bar w_0}{\cal S}^{(0)}=0$ and $\frac{\pa}{\pa t}{\cal
S}^{(0)}=0$,
there are no terms of order $\ka^{-1}$.
In the quasi-classical limit we put
$\pa{\cal S}^{(0)}=\la$.
In the zero order approximation we come to the linear system
of the two-body ECMM (\ref{L1})--(\ref{L3}).
The Baker-Akhiezer function $Y$ takes the form
$$
Y=\Phi e^
{ t\frac{\pa}{\pa \tau_0}{\cal S}^{(0)}}\,.
$$
This passage from the autonomous two-body ECCM to the  Painlev\'{e} VI
equation is an example of the Whitham quantization.
The quasi-classical limit of the full PVI  yields the generalization of
ECMM \cite{In}.

We can consider
the $\SLN$ generalization of the isomonodromy problem.
 The related Lax matrix
has the form
$$
L=P+X\,,~~
P=2\pi i\di\left(\frac{{\bf v}}{1-\mu}-\ka\frac{{\bf u}}{\rho}\right),
$$
$$
X_{jk}=\{x_{\al}\}=(\tau-\bar{\tau}_0)\nu
\phi(u_j-u_k,w).
$$
The multi-component analog of the Painlev\'{e} VI
equation is the monodromy preserving equation
\beq{MC}
\frac{\ka^2d^2u_j}{d\tau^2}=-\frac{\nu^2}{(2\pi i)^2}
\sum_{k< j}^N\pa_uE_2(u_j-u_k|\tau)\,.
\end{equation}
In the quasi-classical limit $\ka\to 0$ we come to the linear problem for
the $N$-body ECMM (\ref{L1})--(\ref{L3}).

\section{Calogero-Moser model and Knizhnik-Zamolodchikov-Bernard equation}
\setcounter{equation}{0}

The Knizhnik-Zamolodchikov-Bernard equation (KZB) is the generalization on a torus of
the Knizhnik-Zamolodchikov equation \cite{KZ} obtained by D.Bernard \cite{B}.
Its solutions are correlation functions of the Wess-Zumino-Witten model on
a the torus with $n$ marked points.
 The KZB equation has the form of
the non-stationer Schr\"{o}dinger equation where the role of times is played
by the module $\tau$ and the position of $n-1$ points.
The classical limit of the KZB equations in general case was considered in
\cite{I}.
 We consider the case
$n=1$. The correlation function $F$ depends on a finite-dimensional
representation $V$ attributing to the marked point.
The KZB equation has the form
\beq{NS}
\left(\ka^{quant}\pa_\tau+
\oh\sum_{j=1}^N\pa^2_{u_j}+\sum_{j>k}\hat{e}_{jk}\hat{e}_{kj}E_2(u_j-u_k;\tau)
\right)F=0\,,
\end{equation}
where $\hat{e}_{kj}$ are generators of the matrix elements $e_{kj}$ in $V$
and $\ka^{quant}=\kappa+N$.
To pass to the classical limit in the KZB equations
we replace the conformal block by its quasi-classical expression
\beq{cb}
F=\exp \frac{\cal F}{\hbar},
\end{equation}
where $\hbar =(\ka^{quant})^{-1}$. Consider the classical limit
$  \ka^{quant} \rar\infty$ and assume that values of the Casimirs
$C^i_a,~(i=1,\ldots,{\rm rank}G,~a=1,\ldots,n)$
corresponding to the
irreducible representations
 also go to infinity. Let all values
$\lim\frac{C_a^i}{\ka^{quant}}$
are finite. It allows to fix the coadjoint orbits in the marked point.
In the classical limit (\ref{NS}) is transformed to the Hamilton-Jacobi
equation for the action ${\cal F}$
$$
\pa_\tau{\cal F}-H^{CM,spin}(\pa_\bfu{\cal F},\bfu)=0\,.
$$
In this way we come to isomonodromy preseving case.

On the critical level $\ka^{quant}=0$ we come to the eigenvalue problem
for the quantum ECMM for the zero eigenvalue. It allows to describe the
wave-functions of the quantum ECMM \cite{EK,FV}.

\bigskip

We summarize the result of
last two sections in the following diagram.
$$
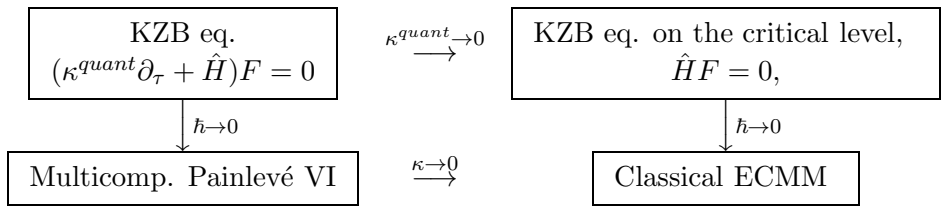
\def\normalbaselines{\baselineskip20pt
       \lineskip3pt    \lineskiplimit3pt}
\def\mapright#1{\smash{
        \mathop{\longrightarrow}\limits^{#1}}}
\def\mapdown#1{\Big\downarrow\rlap
       {$\vcenter{\hbox{$\scriptstyle#1$}}$}}
\begin{array}{ccc}
\fbox{$
\begin{array}{c}
\mbox{KZB eq.}\\
(\ka^{quant}\pa_{\tau}+\hat{H})F=0
\end{array}
$}
  &\mapright{\ka^{quant}\rar0} &
\fbox{$
\begin{array}{c}
\mbox{KZB eq. on the critical level},~\\
\hat{H}F=0,\\
\end{array}
$}
\\
\mapdown{\hbar\rar 0}&    &\mapdown{\hbar\rar 0} \\
\fbox{$
\begin{array}{c}
\mbox{Multicomp. Painlev\'{e} VI}\\
\end{array}
$}
&\mapright{\ka\rar 0} &
\fbox{$
\begin{array}{c}
\mbox{Classical ECMM} ~\\
\end{array}
$}
\end{array}
$$
Here before going from PVI to the classical ECMM we renormalize
the variables and the half-periods according with (\ref{hp}).

\bigskip

{\small {\bf Acknowledgments.}\\
I thank
the Research Institute for Mathematical Science of Kyoto University for the hospitality
where this work was done.
 The work is supported by the grants NSh-1999-2003.2 of the scientific
schools, RFBR-03-02-17554  and CRDF RM1-2545.}

\small{

}
\end{document}